# Diffraction of helical x-rays by optically active achiral crystals

S. W. Lovesey[1,2]

[1]ISIS Facility, STFC, Didcot, Oxfordshire OX11 0QX, UK

[2]Diamond Light Source Ltd, Didcot, Oxfordshire OX11 0DE, UK

**Abstract** Four crystal classes are optically active yet not chiral (non-enantiomorphic). Corresponding Bragg diffraction patterns calculated for circularly polarized (helical) x-rays tuned to an atomic resonance display angular anisotropy in the distribution of electrons at spots not indexed on the chemical structure (defined by a space group derived from Thomson scattering by perfect spheres of charge). Templeton-Templeton scattering, as it is usually called, of helical x-rays is quantified in terms of a chiral signature defined as the partial diffracted intensity hallmarked by x-ray helicity. Our electric dipole - electric dipole (E1-E1) chiral signature for a space group in the optically active crystal class $\bar{4}2m$ ($D_{2d}$) affords a complete interpretation in terms of copper quadrupoles of diffraction data collected on copper metaborate [E. N. Ovchinnikova *et al.*, J. Synchrotron Rad. **28**, 1455 (2021)]. An example of a chiral signature derived from a parity-odd resonance event (E1-E2) is included in our study. It is firmly established that tuning the energy of x-rays to an atomic resonance enhances the intensity of inherently weak Templeton-Templeton scattering and renders space-group forbidden Bragg spots element specific. Our chiral signature as a function of rotation about the reflection vector (azimuthal-angle scan) is specific to position multiplicity, Wyckoff letter and symmetry in a favourable space group.

## I. INTRODUCTION

J. Willard Gibbs, best known for pioneer work on statistical mechanics perhaps, used Maxwell's equations of electrodynamics in a demonstration that certain crystals can be both optically active (rotation of the plane of polarization of light) and achiral (non-enantiomorphic) [1]. Different theory, based on a pseudotensor of rank two, was adopted in subsequent work which established that crystal classes m ($C_s$), mm2 ($C_{2v}$), $\bar{4}$ ($S_4$) and $\bar{4}2m$ ($D_{2d}$) possess the dual properties; see, for example, Refs. [2, 3]. The pseudotensor is certainly zero for materials that possess a centre of inversion symmetry. It can be different from zero for 11 enantiomorphic crystal classes (absence of both a centre and a plane of symmetry), and the named four non-enantiomorphic classes that are identical to their mirror images. Optical experiments on crystal types $\bar{4}2m$ (silver thiogallate) and $\bar{4}$ (cadmium gallium sulphide) by Hobden published between 1967 and 1969 confirmed that non-enantiomorphic crystals can present optical rotation or gyration, as it is sometimes known [4-7]. The crystals show birefringence (double refraction) that overwhelms optical activity. At specific wavelengths, the crystals are accidentally isotropic, however, which allowed Hobden to study optical activity without the difficulties arising from the presence of birefringence.

Coupling of helicity in a beam of x-rays and a crystal chiral axis has been confirmed in diffraction by several enantiomorphic materials [8-13], together with imaging the chirality of domains in racemic twinned Cs Cu Cl$_3$ [14]. Weak Bragg spots not indexed on the chemical crystal structure are created by angular anisotropy in the distribution of atomic electrons. It is often referred to as Templeton-Templeton scattering (T & T) after the pioneers [15, 16]. Signal enhancement is gained by tuning the energy of primary photons to an atomic resonance, which adds specification of an element to the desirable features of the x-ray diffraction technique. To date, there are successful experiments on low-quartz, tellurium and berlinite (Al P O$_4$) with enantiomorphs that belong to trigonal P3$_1$21 (*right*-handed screw) and P3$_2$21 (*left*-handed screw), and resonant ions occupy sites with multiplicity three [8-13]. Diffraction patterns and azimuthal-angle scans (rotation of the diffracting crystal about the axis of the reflection vector) were measured with the energy of primary x-rays was tuned to K-edges (1s) of Si and Al, for example.

We investigate Bragg diffraction of circularly polarized x-rays by crystal classes C$_s$, C$_{2v}$, S$_4$ and D$_{2d}$, and thereby extend our understanding of what can be learned from appropriate experiments. Calculated scattering amplitudes account for the enhancement of Bragg spots by an atomic resonance and a rotation of the crystal about the reflection vector. Use is made of a chiral signature $\Upsilon$ specific to position multiplicity, Wyckoff letter and symmetry in a space group. Our findings imply a universal azimuthal-angle dependence of $\Upsilon$ in diffraction enhanced by electric dipole - electric dipole (E1-E1) and electric dipole - electric quadrupole (E1-E2) absorption events. The signal is absent for the four space groups in the C$_s$ crystal class, and the two space groups in S$_4$. By way of an example, copper and neodymium ions in the one-dimensional type cuprate Nd$_2$ Ba$_4$ Cu$_2$ O$_9$ occupy sites 4f in the tetragonal space group P$\bar{4}$n2 in crystal class D$_{2d}$. Their contributions to a Bragg diffraction pattern are specified by the energy of the chosen atomic resonance, and specific electronic structures create quadrupoles and the chiral signature.

## II. RESONANT SCATTERING

The nature of the electronic ground state accessed in resonant diffraction depends on both the quantum labels of the virtual, intermediate state and the type of the absorption event, whose actual strength may depend on its energy. An E1 event and absorption of hard x-rays (energy E ≈ 5 - 9 keV) at a K-edge can access p-like atomic states, while an E2 event can access d-like states. Levels of enhancement are very small set against factors enjoyed at actinide M$_{4;5}$ absorption edges. Direct access with an E1 event to d-like states is allowed by absorption at L-edges, and an even larger enhancement is expected. It can even be comparable to the intensities of Thomson reflections from the chemical structure. Not all crystal structures satisfy the Laue condition for Bragg diffraction with an x-ray wavelength ≈ (12.4/E) Å and E in units of keV.

Assuming that virtual intermediate states are spherically symmetric, to a good approximation, the x-ray scattering length ≈ {F$_{\mu\eta}$/(E − Δ + iΓ/2)} in the region of the resonance, where Γ is the total width of the resonance at an energy Δ [10, 17-19]. The numerator F$_{\mu\eta}$ is an amplitude, or unit-cell structure factor, for Bragg diffraction in the scattering channel with primary (secondary) polarization η (μ). By convention, σ denotes polarization normal to the plane of scattering, and π denotes polarization within the plane of scattering. Fig. 1 depicts polarization states, wave vectors, and the Bragg condition.

## III. CHIRAL SIGNATURE

Photon and electronic quantities in the scattering amplitude are partitioned in a generalized scalar product $F_{\mu\eta} = \{\mathbf{X}^K \bullet \langle \mathbf{O}^K \rangle\}$, with implied sums on rank K and associated projections Q in the interval $-K \leq Q \leq K$ [18,19]. Angular brackets about the atomic tensor operator $O^K_Q$ in an electronic multipole $\langle O^K_Q \rangle$ denote its time-average, or expectation value. Selection rules on K and Q for the multipole imposed by symmetry of the site used by the resonant ion are evidently duplicated in the x-ray factor $\mathbf{X}^K$. The latter is specific to a resonant event. One finds, $\mathbf{X}^K$ is independent of photon wave vectors for an E1-E1 event (K = 0 - 2) but this is not so for E1-E2 (K = 1 - 3) and E2-E2 (K = 0 - 4) absorption events. All information on x-ray factors needed here is found in Refs. [18, 19]. Electronic multipoles can be calculated using standard tools of atomic physics given a suitable wavefunction [9]. Alternatively, multipoles can be estimated from a tried and tested simulation program of electronic structure [10]. The complex conjugate of an atomic multipole $\langle O^K_Q \rangle^* = (-1)^Q \langle O^K_{-Q} \rangle$, with a phase convention $\langle O^K_Q \rangle = [\langle O^K_Q \rangle' + i\langle O^K_Q \rangle'']$ for real and imaginary parts labelled by single and double primes, respectively.

Henceforth, we adopt a shorthand $(\mu\eta)$ for the scattering amplitude $F_{\mu\eta}$. Scattered intensity picked out by circular polarization in the primary photon beam $= P_2 \Upsilon$,

$$\Upsilon = \{(\sigma'\pi)^*(\sigma'\sigma) + (\pi'\pi)^*(\pi'\sigma)\}'', \qquad (1)$$

and the Stokes parameter $P_2$ (a purely real pseudoscalar) measures helicity in the primary x-ray beam; cf. Eq. (12) in Ref. [10]. Since intensity is a scalar quantity, $\Upsilon$ and $P_2$ must possess identical discrete symmetries, specifically, both scalars are time-even and parity-odd. The signature $\Upsilon$ is extracted from observed intensities by subtraction of intensities measured with opposite handed primary x-rays, namely, $\pm P_2$. Intensity of a Bragg spot in the rotated channel of polarization is proportional to $|(\pi'\sigma)|^2$, and likewise for unrotated channels of polarization.

We use $\Psi^K_Q = [\exp(i\boldsymbol{\kappa} \bullet \mathbf{d}) \langle O^K_Q \rangle_\mathbf{d}]$ for an electronic structure factor, where the reflection vector $\boldsymbol{\kappa}$ is defined by integer Miller indices ($h$, $k$, $l$), and the implied sum in $\Psi^K_Q$ is over all sites $\mathbf{d}$ in a unit cell used by resonant ions. Construction of $\Psi^K_Q$ requires symmetry operators for a space group together with the symmetry of occupied sites. All necessary information is available on the Bilbao server [20]. The four amplitudes required in the chiral signature defined in Eq. (1) are derived from $\Psi^K_Q$ and universal expressions for E1-E1 and E1-E2 scattering amplitudes [19].

## IV. OPTICALLY ACTIVE ACHIRAL SPACE GROUPS

Calculations return null chiral signatures for crystal classes $C_s$ and $S_4$. For the two remaining optically active achiral space groups our goal is to raise awareness of the virtue of the chiral signature. Many real samples have been identified [21], and likely more wait to be discovered, and it is not prudent to attempt construction of a catalogue of materials.

## A. $C_{2v}$

There are 22 space groups in the orthorhombic crystal class mm2 ($C_{2v}$), with cell lengths $a \neq b \neq c$. Space group Pca$2_1$ (No. 29) is chosen as an opening illustration because it consists of sites 4a alone. Sites have no symmetry. In consequence, projections Q in the interval $-K \leq Q \leq K$ for a multipole $\langle O^K_Q \rangle$ are unrestricted. The electronic structure factor is,

$$\Psi^K_Q(4a) = \langle O^K_Q \rangle [\exp(i\delta) + (-1)^{Q+l} \exp(-i\delta)] \quad \text{(No. 29)}$$

$$+ \sigma_\pi \langle O^K_Q \rangle^* (-1)^{K+h} [\exp(i\chi) + (-1)^{Q+l} \exp(-i\chi)], \quad (2)$$

where $\delta = 2\pi(xh + yk)$, $\chi = 2\pi(xh - yk)$ and x and y are general coordinates. The parity signature $\sigma_\pi = +1$ ($-1$) for a parity-even (parity-odd) absorption event, e.g., E1-E1 (E1-E2). Examination of $\Psi^K_0(4a)$ with K even and $\sigma_\pi = +1$ reveals that reflections ($h$, 0, 0) and (0, $k$, $l$) with Miller indices $h$ and $l$ odd, respectively, are not indexed on the crystal structure, i.e., $\Psi^K_0(4a)$ is zero for said conditions. T & T scattering at such Bragg spots is created by non-diagonal multipoles with projections $|Q| > 0$.

We proceed with a study of ($h$, 0, 0) and $h$ odd utilizing the E1-E1 event. Multipoles with $\sigma_\pi = +1$ are denoted $\langle T^K_Q \rangle$. Let the two quantities A $= -4 \sin(\varphi_h) \langle T^2_{+1} \rangle''$ and B $= 4 i \cos(\varphi_h) \langle T^2_{+2} \rangle''$ in which $\delta = \chi = \varphi_h = 2\pi xh$. Scattering amplitudes are [19],

$$(\sigma'\sigma) = -i \sin(2\psi) A, \quad (\pi'\pi) = \sin^2(\theta) (\sigma'\sigma), \quad (3)$$

$$(\pi'\sigma) = -i \sin(\theta) \cos(2\psi) A + i \cos(\theta) \sin(\psi) B, \quad (2m + 1, 0, 0; \text{No. 29})$$

and ($\sigma'\pi$) is derived from ($\pi'\sigma$) by a change in sign to A. The Bragg angle $\theta$ is defined in Fig. 1. Note that amplitudes ($\sigma'\sigma$) and ($\pi'\pi$) are purely imaginary, while ($\sigma'\pi$) and ($\pi'\sigma$) are complex. The azimuthal angle $\psi$ measures rotation of the crystal about the reflection vector, and the crystal b-axis is in the plane of scattering for $\psi = 0$. It follows that the chiral signature of the Bragg spot ($h$, 0, 0) enhanced by an E1-E1 absorption event is,

$$\Upsilon_+(4a) = -8 \Phi(\varphi_h, \psi) \langle T^2_{+1} \rangle'' \langle T^2_{+2} \rangle'', \quad (2m + 1, 0, 0; \text{No. 29}) \quad (4)$$

where,

$$\Phi(\varphi_h, \psi) = \cos^3(\theta) \sin(2\varphi_h) \sin(\psi) \sin(2\psi). \quad (5)$$

The azimuthal-angle dependence captured in $\Phi(\varphi_h, \psi)$ appears to be universal to E1-E1 and E1-E2 absorption events. The two multipoles that create $\Upsilon$ have angular characters (yz) and (xy). Dependence of $\Upsilon$ on $\varphi_h = 2\pi xh$ will likely generate optimal chiral signatures at particular Bragg spots. The subscript on $\Upsilon_+(4a)$ denotes the fact that $\sigma_\pi = +1$.

The fundamental structure of $\Upsilon_-(4a)$ for an E1-E2 event is similar to $\Upsilon_+(4a)$. The number of multipoles engaged is the defining difference, and it arises from the fact that polar multipoles $\langle U^K_Q \rangle$ have ranks K = 1, 2, and 3 and projections Q are unrestricted by site symmetry. Scattering amplitudes in unrotated channels of polarization are purely imaginary. Amplitudes in rotated channels are complex, and their real parts allow $\Upsilon_-(4a)$ different from zero. For E1-E2 we find,

$$\Upsilon_-(4a) = (16/(15\sqrt{5})) \sin^2(\theta)\, \Phi(\varphi_h, \psi)\, [3\, \langle U^1_{+1}\rangle'' + 2\sqrt{5}\, \langle U^2_{+1}\rangle' - \langle U^3_{+1}\rangle'' + \sqrt{15}\, \langle U^3_{+3}\rangle'']$$

$$\times [\sqrt{6}\, \langle U^2_0\rangle + \langle U^2_{+2}\rangle' + \sqrt{2}\, \langle U^3_{+2}\rangle'']. \quad (2m+1, 0, 0;\ \text{No. 29}) \quad (6)$$

Notably, $\Upsilon_+(4a)$ and $\Upsilon_-(4a)$ have exactly the same dependence on the azimuthal angle. Also, both chiral signatures are created by the interference of multipoles with even and odd projections.

Space group Pna2$_1$ (No. 33) likewise comprises sites 4a alone that have no symmetry. Reflections (0, k, l) with k + l odd are space-group forbidden, and $\Upsilon_+(4a) \propto [\langle T^2_{+1}\rangle'\, \langle T^2_{+2}\rangle'']$ for an E1-E1 absorption event.

A more typical structure comprises sites with more than one multiplicity. There are four sites with multiplicity two in space group Pcc2 (No. 27). All have site symmetry 2$_z$. The latter restricts projections to Q = ± 2n, including Q = 0. Diagonal components (Q = 0) of the electronic structure factor, by definition, are zero at space group forbidden reflections. Quadrupoles in T & T scattering are $\langle T^2_{\pm 2}\rangle$ at sites with 2$_z$ symmetry. In consequence, the chiral signature is zero for such high symmetry sites since it relies on interference between multipoles that differ by projections or ranks, or both; cf. Eq. (6).

Remaining positions 4e in Pcc2 have no symmetry. Reflections (h, 0, l) and (0, k, l) with l odd are forbidden. Consider the Bragg spot (h, 0, l) first. Scattering amplitudes are functions of the two quantities,

$$A = 4\, [p \sin(\varphi_h)\, \langle T^2_{+1}\rangle'' - i\, r \cos(\varphi_h)\, \langle T^2_{+2}\rangle''], \quad (7)$$

$$B = -4\, [p \cos(\varphi_h)\, \langle T^2_{+1}\rangle'' + i\, r \sin(\varphi_h)\, \langle T^2_{+2}\rangle'']. \quad (h, 0, 2m+1;\ \text{No. 27})$$

Scattering amplitudes are derived from Eq. (3) on using these expressions. In Eq. (7), (p, 0, r) is a unit vector parallel to the reflection vector, i.e., p ∝ h and r ∝ (al/c). With $\psi = 0$ and (h, 0, 2m + 1) the crystal b-axis is in the plane of scattering, and for (0, k, 2m + 1) the a-axis is in the plane.

Chiral signatures of the Bragg spots (h, 0, l) and (0, k, l) with l odd are,

$$\Upsilon_+(4e) = -8\, \Phi(\varphi_h, \psi)\, \langle T^2_{+1}\rangle''\, \langle T^2_{+2}\rangle'', \quad (h, 0, 2m+1;\ \text{No. 27})$$

$$\Upsilon_+(4e) = 8\, \Phi(\varphi_k, \psi)\, \langle T^2_{+1}\rangle'\, \langle T^2_{+2}\rangle''. \quad (0, k, 2m+1;\ \text{No. 27}) \quad (8)$$

Evidently, $\Upsilon_+(4e) = 0$ for (0, 0, 2m + 1). Components of $\langle T^2_{+1}\rangle = [\langle T^2_{+1}\rangle' + i\langle T^2_{+1}\rangle'']$ are different at the two Bragg spots.

## B. $D_{2d}$

Copper and neodymium ions in $Nd_2 Ba_4 Cu_2 O_9$ occupy sites 4f in the tetragonal space group $P\bar{4}n2$ (No. 118) [22], with cell lengths $a = b \neq c$. Multipoles are unchanged by a dyad $2_{-xy}$, and Thomson diffraction is forbidden for reflections of the type $(h, 0, 0)$ with $h$ odd. General extinction rules for $P\bar{4}n2$ are readily derived from the electronic structure factor,

$$\Psi^K_Q(4f) = \langle O^K_Q \rangle [\exp(i\chi) + (-1)^Q \exp(-i\chi)] \quad \text{(No. 118)}$$

$$+ \sigma_\pi \langle O^K_{-Q} \rangle (-1)^{K+Q} (-1)^{h+k+l} [\exp(i\delta) + (-1)^Q \exp(-i\delta)], \quad (9)$$

with $x = y$ in the spatial angles $\chi$ and $\delta$. Multipoles are obliged by site symmetry to satisfy $\langle O^K_Q \rangle = (-1)^K \exp(-i\pi Q/2) \langle O^K_{-Q} \rangle$.

Scattering amplitudes and the chiral signature for an E1-E1 event are,

$$(\sigma'\sigma) = 4i \sin(\varphi) \sin(2\psi) \langle T^2_{+1} \rangle', \quad (\pi'\pi) = \sin^2(\theta) (\sigma'\sigma),$$

$$(\pi'\sigma)' = (\sigma'\pi)' = -4 \cos(\varphi) \cos(\theta) \sin(\psi) \langle T^2_{+2} \rangle'', \quad (2m+1, 0, 0; \text{No. 118})$$

$$\Upsilon_+(4f) = -8 \Phi(\varphi_h, \psi) \langle T^2_{+1} \rangle' \langle T^2_{+2} \rangle'', \quad (10)$$

with $\varphi_h = 2\pi x h$. General coordinates for ions using 4f are significantly different and mean optimal Bragg spots occur at different Miller indices $h$ ($Nd_2 Ba_4 Cu_2 O_9$, $x \approx 0.388$ (Nd) and 0.101 (Cu) [22]). Quadrupoles satisfy $\langle T^2_{+1} \rangle' = \langle T^2_{+1} \rangle''$.

As a second example of the crystal class type, consider space group $I\bar{4}2d$ (No. 122) with the condition $h + k + l$ even imposed on Miller indices by body centring. Compounds using this space group include $Ag Ga S_2$, $Cu Fe S_2$, and $Cu B_2 O_4$. Ions occupy sites with multiplicities 4 and 8. Sites 4a, 4b are not of interest in the study in hand because they do not contribute T & T diffraction in E1-E1. Not so for sites 8d with symmetry $2_x$ used by $S^{2-}$ ions in silver thiogallate and chalcopyrite, and $Cu^{2+}$ ions in copper metaborate [23]. Reflections $(h, h, 0)$ with $h$ odd are space-group forbidden, for example. The corresponding chiral signature $\Upsilon_+(8d)$ is the same as in Eq. (10) on replacing quadrupoles therein by $\langle T^2_{+1} \rangle''$ and $\langle T^2_{+2} \rangle'$, and multiplication by a factor $-2\sqrt{2}$. The fractional coordinate $x \approx 0.082$ for copper ions in $Cu B_2 O_4$ [23]. Our result $\Upsilon_+(8d)$ accounts for an azimuthal scan on the Bragg spot $h = 1$, and it specifies the nature of Cu quadrupoles exposed by helicity in the primary beam of x-rays [24].

## V. DISCUSSION

An answer to the intriguing question as to whether optically active achiral crystals possess a chemical structure that interacts with helicity in a beam of x-rays is found in work reported by Ovchinnikova *et al.* [24]. The authors measured a partial intensity, defined as the difference between the intensity of Bragg spots observed with left and right-handed x-ray, for copper metaborate with the energy of x-rays tuned to the copper K-edge ($E \approx 8991$ eV, $1s \rightarrow 4p$). The material is described by one of 12 space groups in the optically active non-enantiomorphic crystal class $D_{2d}$.

Our calculations using $C_{2v}$ and $D_{2d}$ of the partial intensity, a chiral signature denoted by ϒ, imply that it has a universal structure for diffraction enhanced by an electric dipole - electric dipole (E1-E1) absorption event. As a function of rotation through an azimuthal angle ψ about the reflection vector, ϒ ∝ sin(ψ) sin(2ψ) and, thus, zero for ψ multiples of 90°, and its size and sign depend on Miller indices. As in Templeton-Templeton scattering, quadrupoles make up the electronic content of ϒ, and we identify the specific quadrupoles observed in the experiment reported on Cu $B_2$ $O_4$ [24]. Multipoles are perfectly defined atomic quantities that can be calculated using a suitable electronic wavefunction, or calculated using a simulation of electronic structure [18, 25].

Small intensity was measured ≈ 10 eV below the main intensity at an energy E ≈ 8991 eV in the copper absorption spectrum, at which Bragg diffraction was performed [24]. A parity-odd E1-E2 absorption event is one candidate mechanism for the blip in intensity. We report a calculation of the chiral signature for E1-E2 and find it is essentially the same as for an E1-E1 event. Specifically, the azimuthal-angle dependence is the same for both absorption events. While ϒ(E1-E1) is a product of two parity-even quadrupoles with different angular anisotropies ϒ(E1-E2) can contain polar multipoles with different ranks and projections, as in the example we report, Eq. (6). We have not found ϒ(E1-E1) different from zero for chemical structures that belong to $C_s$ and $S_4$ crystal classes.

A few materials that belong to crystal classes $C_{2v}$ and $D_{2d}$ are listed in Table I. Our chiral signature defined in Eq. (1) can be different from zero for ions using the sites mentioned. Its specific form is deduced from results presented in the main text, and always it is proportional to $\Phi(\varphi, \psi)$ in Eq. (5). In the example of the tellurium sulfate [26], the signature ϒ(4b) for space-group forbidden spots $(2m + 1, 0, 0)$ is given by Eq. (4) apart from an overall minus sign. With a cell length $a \approx 8.880$ Å, recovered from powder neutron diffraction patterns, the Laue condition for Bragg diffraction is not satisfied at the oxygen K-edge. Absorption at the tellurium $L_3$ edge (E ≈ 4.345 keV) gives access to Bragg spots with Miller indices $h = 1, 3$ & 5, and many more spots are accessible with absorption at the K-edge (E ≈ 31.817 keV). From Eq. (4), ϒ(4b) ∝ sin(4πx$h$) with x ≈ 0.301 for tellurium, and sin(4πx$h$) ≈ − 0.935 (+ 0.980) for $h = 3$ (7), while this spatial phase factor ≈ + 0.082 for $h = 5$. Resonance enhanced Bragg diffraction by oxygen ions can be realized with crystals of Nd Os $O_4$ since the cell length is much larger $a \approx 14.859$ Å [28]. Regarding the other two elements in this compound, x-rays tuned to the $M_3$ and $M_4$ edges likely yield strong enhancements, with the $M_4$-edge at E ≈ 1.000 keV (E ≈ 2.033 keV) for Nd (Os).

**ACKNOWLEDGEMENT** Professor G. van der Laan prepared Fig. 1, and Dr Y. Tanaka commented on the paper in its making.

----------------------------------------------------------------------------------------------------

**TABLE I.** Representative examples of optically active achiral materials in crystal classes $C_{2v}$ (space groups Nos. 25-46, §IV. A) and $D_{2d}$ (Nos. 111-122, §IV. B). Our chiral signature $\Upsilon$ defined in Eq. (1) can be different from zero for ions in the cited sites.

$Te_2 O_3 (S O_4)$, $Pmn2_1$ (No. 31) sites 4b; $Te^{4+}$, $O^{2-}$ [26]

$Na_3 W O_3 N$, $Pmn2_1$ (No. 31) sites 4b; $Na^{1+}$, $O^{2-}$, $N^{3-}$ [27]

$Nd Os O_4$, $Pna2_1$ (No. 33) sites 4a; $Nd^{3+}$, $Os^{5+}$, $O^{2-}$ [28]

$Rb_2 Cd Br_2 I_2$, $Ama2$ (No. 40) sites 8c; I [29]

$Ba Cu_2 Sn Se_4$, $Ama2$ (No. 40) sites 8c; Cu, Se [30]

$Nd_2 Ba_4 Cu_2 O_9$, $P\bar{4}n2$ (No. 118) sites 4f; $Nd^{3+}$, $Cu^{2+}$, $O^{2-}$ [22]

$Cu B_2 O_4$, $I\bar{4}2d$ (No. 122) sites 8d; $Cu^{2+}$ [23]

$Ag Ga S_2$, $I\bar{4}2d$ (No. 122) sites 8d; $S^{2-}$ [31]

----------------------------------------------------------------------------------------------------

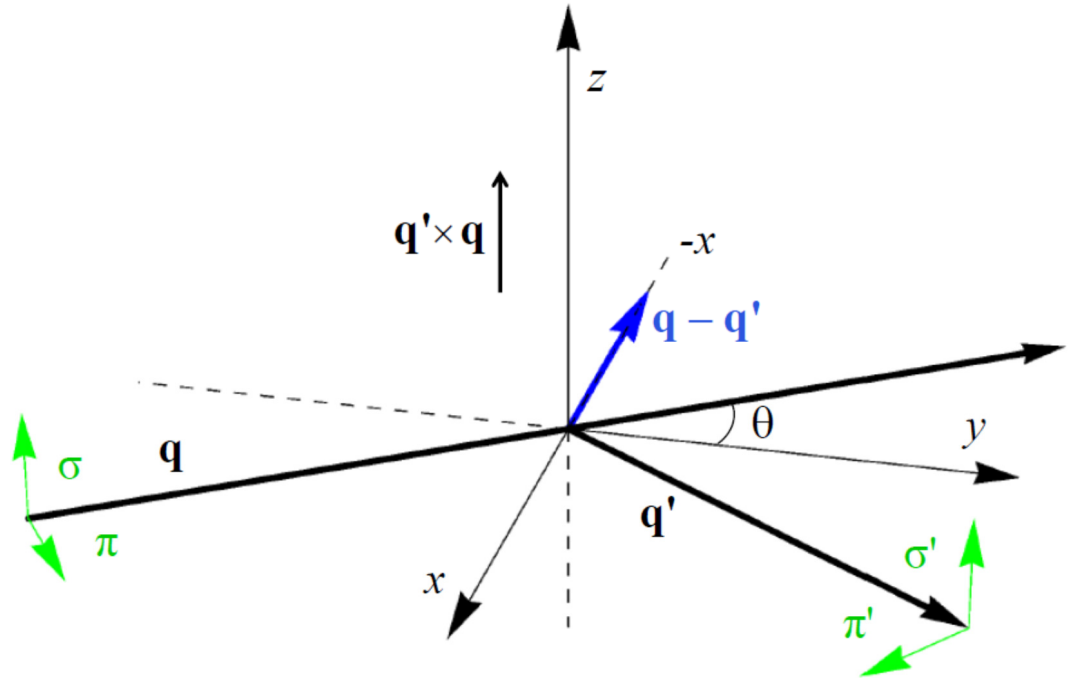

FIG. 1. Primary ($\sigma$, $\pi$) and secondary ($\sigma'$, $\pi'$) states of polarization. Corresponding wavevectors **q** and **q**′ subtend an angle $2\theta$, and the reflection vector $\boldsymbol{\kappa} = \mathbf{q} - \mathbf{q}'$.